\documentclass[11pt,a4paper]{article}
\pdfoutput=1
\usepackage{jcappub}

\usepackage[T1]{fontenc}
\usepackage{graphicx}
\usepackage{amsfonts,amsmath,amssymb,tensor}
\usepackage{mathtools}
\usepackage{braket}
\usepackage{bm}
\usepackage{enumerate}
\usepackage{ulem}
\usepackage{comment}

\usepackage[dvipsnames]{xcolor}

\begin{document}

\title{Formation of primordial black holes through Q-balls}
\author[a]{Shinta Kasuya,}
\author[b,c]{Masahiro Kawasaki,}
\author[d,c]{Alexander Kusenko,}
\author[b]{Shunsuke Neda}

\affiliation[a]{Physics Division, Faculty of Science, Kanagawa University, Kanagawa 221-8686, Japan}
\affiliation[b]{ICRR, University of Tokyo, Kashiwa, 277-8582, Japan}
\affiliation[c]{Kavli IPMU (WPI), UTIAS, University of Tokyo, Kashiwa, 277-8583, Japan}
\affiliation[d]{Department of Physics and Astronomy, University of California, Los Angeles, CA 90095-1547, USA}

\abstract{%
We study the primordial black hole (PBH) formation from Q-balls that are non-topological solitons in scalar field theories.
We develop a formula for calculating the density perturbations from the Q-ball charge distribution.
We also re-examine the condition for the PBH formation in the matter-dominated era and show that the previously derived formula for super-horizon density fluctuations can be applied to the sub-horizon density perturbations.
As an example, we consider the Q-balls in the case of gauge-mediated supersymmetry (SUSY) breaking, whose charge distribution was obtained by the lattice simulation.  We find that the density perturbations are large enough to produce a significant number of PBHs with mass $10^{-15}\,M_\odot -5\times 10^{-12}\, M_\odot$, which can explain all the dark matter in the universe. In the context of supersymmetry, this mass range corresponds to the SUSY breaking scale $\sim 10^6$~GeV, which is consistent with the SUSY particle masses $\sim 10$~TeV.

}


\maketitle

\section{Introduction}
\label{sec: intro}

Q-balls are non-topological solitons that appear in a complex scalar field theory with a global $U(1)$ symmetry~\cite{Coleman:1985ki}.
Q-balls are classically stable due to $U(1)$ charge conservation.
In the early universe, Q-balls are formed when the scalar field starts oscillating in a potential flatter than the quadratic one.
Such a flat potential is naturally realized in the framework of supersymmetric (SUSY) theories.
In particular, Q-balls are produced by the Affleck-Dine mechanism~\cite{Affleck:1984fy,Dine:1995kz}, which utilizes the flat directions in the minimal SUSY standard model~\cite{Kusenko:1997zq,Kusenko:1997si,Dvali:1997qv,Enqvist:1997si,Kasuya:1999wu}.

When the Q-balls are formed from an oscillating scalar field with a large amplitude $\phi_\mathrm{osc}$, the formed Q-balls have large charges $Q\sim (\phi_\mathrm{osc}/\mu)^2$ or $Q\sim (\phi_\mathrm{osc}/\mu)^4$, depending on the shape of the potential~\cite{Dvali:1997qv}, and masses $M_Q\sim \phi_\mathrm{osc}^2/\mu$ or $M_Q\sim \phi_\mathrm{osc}^3/\mu^2$,  where $\mu (\, \ll \phi_\mathrm{osc})$ is the mass scale of the potential.
In this case,  their number is small and thus fluctuates significantly due to Poisson fluctuations, which lead to large density fluctuations of the Q-balls. 
When the Q-balls dominate the density of the universe, their large density fluctuations can generate primordial black holes (PBHs) as pointed out in~\cite{Cotner:2016cvr, Cotner:2017tir,Cotner:2019ykd}.
In the previous studies, however, it was assumed that the Q-balls have a monochromatic charge or mass, while the lattice simulation shows that Q-balls with a wide range of charges are produced~\cite{Hiramatsu:2010dx,Kasuya:2010vq}.
Furthermore, when the Q-ball density fluctuations are produced inside the horizon, the condition for PBH formation is not well understood.

In this paper, therefore, we revisit the primordial black hole (PBH) formation by Q-balls proposed in~\cite{Cotner:2016cvr,Cotner:2017tir,Cotner:2019ykd} (see also \cite{Cotner:2018vug}), and refine and develop the formalism for estimating the density perturbations of the Q-balls and the formation rate of PBHs.
The main improvements that we make to the previous studies are as follows:
\begin{itemize}
    \item We develop the formalism that calculates the density perturbations of the Q-balls with an arbitrary charge distribution.
    This enables us to estimate the Q-ball density perturbations from the lattice simulations that show a broad charge distribution of the formed Q-balls.
    \item We derive the formula for estimating the fraction of overdensity regions that collapse into black holes when the density perturbations are generated inside the horizon in the matter-dominated universe.
\end{itemize}
It is found that the previous formula for the PBH formation, which is derived (implicitly) assuming super-horizon scale density perturbations, can also be applied to the sub-horizon case.
We estimate the abundance of PBHs using the above formula and the result of the recent simulation for gauge-mediation type Q-balls~\cite{Kasuya:2025nix}, and find that the PBHs with mass $10^{-15}\,M_\odot -5\times 10^{-12}\, M_\odot$ can account for all the dark matter of the universe.
Furthermore, since the Q-balls most naturally appear in the framework of SUSY, we consider the Q-ball scenario for the PBH formation in the gauge-mediation SUSY breaking models, and obtain a relation between the SUSY breaking scale and the dark matter PBH mass.

The rest of this paper is organized as follows.
In Sec.~\ref{sec:Qball_perturbation}, we derive the formula for estimating the density perturbations of the Q-balls with a charge distribution.
The PBH formation rate in a matter-dominated universe is re-examined for sub-horizon density perturbations in Sec.~\ref{sec:PBH_formation}.
We estimate the PBH abundance and mass using the result of the lattice simulation in Sec.~\ref{sec:PBH_formation_Qball} and consider the constraint on the present scenario in Sec.~\ref{sec:constraint}.
The Q-balls in supersymmetric models are studied in Sec.~\ref{sec:SUSY_model}.
Our results are summarized and discussed in Sec.~\ref{sec: summary and discussion}.

\section{Density perturbations of Q-balls}
\label{sec:Qball_perturbation}

In this section, we discuss density perturbations induced by the Q-ball distribution.
Suppose that Q-balls in a volume $V$ have the charge distribution $N(Q)$ in the range $[Q_\mathrm{min}, Q_\mathrm{max}]$.
Let us divide this charge range into $n$ small bins $\{ [Q_1, Q_2], [Q_2,Q_3], \ldots [Q_{n1},Q_{n+1}]\}$ ($Q_1 =Q_\mathrm{min}$ and $Q_{n+1} = Q_\mathrm{max}$).
If we take a sufficiently large number of bins, the total energy $E$ is given by
\begin{equation}
    \label{eq:Q-ball-_total_energy}
    E = \sum_{i=1}^{n} M_Q(Q_i) N(Q_i)\Delta Q_i,
\end{equation}
where $\Delta Q_i = Q_{i+1}-Q_i$, and $M_Q(Q)$ is  the mass of Q-balls with charge $Q$.
The mass of Q-balls depends on the potential of the scalar field and is written as
\begin{equation}
    \label{eq:Q-ball_mass}
    M_Q(Q) = \mu Q^p \quad
    \begin{cases}
        p = 3/4 &  \textrm{(gauge-mediation type)} \\[0.4em]
        p=1     &   \textrm{(gravity-mediation type)} ,
    \end{cases}
\end{equation}
where $\mu$ is a mass scale.
The average energy is then given by
\begin{equation}
    \label{eq:Q-ball_ave_energy}
    \langle E \rangle = \sum_{i=1}^{n} M_Q(Q_i)\langle  N(Q_i)\rangle \Delta Q_i
    = \sum_{i=1}^{n} M_Q(Q_i)\bar{N}(Q_i) \Delta Q_i,
\end{equation}
where $\langle \cdots\rangle$ denotes the ensemble average and $\bar{N}(Q) = \langle N(Q)\rangle$.
From Eqs.~\eqref{eq:Q-ball-_total_energy} and \eqref{eq:Q-ball_ave_energy}, the fluctuation of the total energy is written as
\begin{equation}
    \delta E = E-\langle E \rangle = \sum_{i=1}^{n} M_Q(Q_i) \delta N(Q_i)\Delta Q_i ,
\end{equation}
where $\delta N(Q) = N(Q)-\bar{N}(Q)$.
When we assume that each $N(Q_i)$ follows the independent Poisson statistics, the variance of the energy fluctuation is given by
\begin{align}
    \langle (\delta E )^2\rangle  
    &=  \sum_{i=1}^{n} M_Q^2(Q_i) \left\langle (\delta N(Q_i)\Delta Q_i)^2\right\rangle 
    \nonumber \\
    & = \sum_{i=1}^{n} M_Q^2(Q_i) \bar{N}(Q_i)\Delta Q_i .
    \label{eq:Q-ball_fluctuation}
\end{align}
Here we have used $\langle (\delta N(Q_i)\Delta Q_i)^2\rangle = \bar{N}(Q_i)\Delta Q_i$ in the Poison statistics since $N(Q_i)\Delta Q_i$ and $\delta N(Q_i)\Delta Q_i$ are the Q-ball number and its fluctuation in the $[Q_i, Q_{i+1}]$-bin.
Eq.~\eqref{eq:Q-ball_fluctuation} implies that the variance is calculated by the following integral:
\begin{equation}
    \label{eq:Q-ball_energy_fluc}
    \langle (\delta E )^2\rangle  = \int_{Q_\mathrm{min}}^{Q_\mathrm{max}} dQ\,
    M_Q^2(Q) \bar{N}(Q) .
\end{equation}

Suppose that the average charge distribution of Q-balls is given by the following power law with a cutoff:
\begin{equation}
    \label{eq:Q-ball_charge_dist}
    \bar{N}(Q) = 
        \begin{cases}
            \mathcal{N}_Q V \left(\frac{Q}{Q_*}\right)^{\alpha_Q}\frac{1}{Q}
            & \qquad (Q < Q_*) \\
             ~~0 
             & \qquad (Q > Q_*) ~,
        \end{cases}
\end{equation}
where $\mathcal{N}_Q$ is the normalization factor.
Using the distribution~\eqref{eq:Q-ball_charge_dist}, the average energy density of the Q-balls is calculated as 
\begin{align}
    \bar{\rho}_Q  =\frac{\langle E \rangle}{V} 
     &= \frac{1}{V}\int dQ\, M_Q(Q)\bar{N}(Q) ,\\
     &=  \mathcal{N}_Q \int^{Q_*}dQ\, 
     M_Q(Q)\left(\frac{Q}{Q_*}\right)^{\alpha_Q}\frac{1}{Q} 
    = \mathcal{N}_Q  \frac{M_Q(Q_*)}{\alpha_Q + p} .
\end{align}
Here we assume $\alpha_Q + p >0$.
From Eqs.~\eqref{eq:Q-ball_energy_fluc} and \eqref{eq:Q-ball_charge_dist} we obtain the variance of the energy density fluctuations as
\begin{align}
    \langle (\delta\rho_Q)^2\rangle  \equiv 
    \frac{\langle (\delta E )^2\rangle}{V^2}
    & = \frac{1}{V^2} \int dQ M_Q^2 \bar{N}(Q) , \\
    &= \frac{\mathcal{N}_Q}{V}\int^{Q_*}dQ\, 
     M_Q^2(Q)\left(\frac{Q}{Q_*}\right)^{\alpha_Q}\frac{1}{Q} 
    = \frac{\mathcal{N}_Q}{V} \frac{M_Q^2(Q_*)}{\alpha_Q +2p} .
\end{align}
Therefore, the square root of the variance of the density perturbation $\sigma_Q$ in a volume $V$ is given by
\begin{equation}
    \label{eq:Q-ball_sigma}
    \sigma_Q (V) = \sqrt{\frac{\langle (\delta \rho_Q )^2\rangle }{\bar{\rho}_Q^2}}
    = \frac{1}{\sqrt{\mathcal{N}_Q V}}\frac{\alpha_Q +p}{\sqrt{\alpha_Q +2p}}.
\end{equation}

Let us derive the power spectrum of the Q-ball density perturbations.
To this end, we introduce the density perturbation of Q-balls at $\bm{x}$ and its Fourier transform as $\delta_Q(\bm{x})$ and $\delta_Q(\bm{k})$, respectively.
Then, the density perturbation in a volume $V (=4\pi R^3/3) $ around $\bm{x}$ is calculated as
\begin{equation}
    \delta_Q (\bm{x},V) = \int d^3 x' \delta_Q(\bm{x}') W\left(|\bm{x}'-\bm{x}|,R=(3V/(4\pi))^{1/3}\right),
\end{equation}
where $W(\bm{x},V)$ is the top-hat window function,
\begin{equation}
    W(\bm{x},R) = 
    \begin{cases}
        \frac{1} {V} &  ( |\bm{x}| \le R ) \\[0.4em]
        0   & ( |\bm{x}| > R ) .
    \end{cases}
\end{equation}
The Fourier transformation $\delta_Q(\bm{k},V)$ is given by
\begin{equation}
    \delta_Q (\bm{k}, V) = \frac{1}{(2\pi)^3}\int d^3x\, 
    \delta_Q(\bm{x},V) e^{i\bm{k}\cdot\bm{x}}
    =\delta_Q(\bm{k})\,W(\bm{k},R),
\end{equation}
where $W(\bm{k},R)$ are the Fourier modes of $W(\bm{x},R)$ and is given by
\begin{equation}
    W(\bm{k},R) = 3 \frac{\sin (k R) - kR\cos(kR)}{(kR)^3}.
\end{equation}
We define the power spectrum of the density perturbations $P_Q(k)$ as
\begin{equation}
    \langle \delta_Q(\bm{k})\delta_Q(\bm{k}') \rangle
    = (2\pi)^3 \delta^{(3)}(\bm{k}+\bm{k}') P_Q(k) .
\end{equation}
The variance $\sigma_Q^2$ is then calculated as
\begin{align}
    \sigma_Q^2(V) & = \langle \delta_Q(\bm{x},V)^2\rangle  
    \nonumber \\
    & = \left\langle \int \frac{d^3k}{(2\pi)^3}\frac{d^3q}{(2\pi)^3} e^{i\bm{k}\cdot\bm{x}}e^{i\bm{q}\cdot\bm{x}}
    \delta_Q(\bm{k})\,W(\bm{k},R)\delta_Q(\bm{q})\,W(\bm{q},R)\right\rangle
    \nonumber \\
    & = \frac{1}{(2\pi)^3}\int d^3k P_Q(k) \,W(\bm{k},R)^2.
\end{align}
Since $P_Q(k)$ is independent of $k$ for the Poisson distribution, the above equation is written as
\begin{equation}
    \sigma_Q^2 (V) =  \frac{1}{V}P_Q (k),
\end{equation}
where we have used 
\begin{equation}
    \int dx \, x^2 
    \left(3 \frac{\sin x - x\cos x}{x^3}\right)^2 = \frac{\pi}{6} .
\end{equation}
Using Eq.~\eqref{eq:Q-ball_sigma}, we obtain
\begin{equation}
    \label{eq:Q-ball_power_spec}
    P_Q(k) = \frac{1}{\mathcal{N}_Q }\frac{(\alpha_Q +p)^2}{\alpha_Q +2p}.
\end{equation}

Here we remark on the effect of energy conservation of Q-balls.
All charge in the universe is presumed to be confined inside Q-balls.
For gravity-mediation type Q-balls, since the energy of a Q-ball is proportional to its charge, the charge conservation implies the energy conservation.
Thus, the total energy of the Q-balls in a volume larger than the horizon should be conserved, which leads to $\delta \rho_Q =0$ at super-horizon scales.  
The suppression of the power spectrum $P_Q(k)$ due to energy conservation is obtained in~\cite{Kawasaki:2020jnw} as
\begin{equation}
    \label{eq:suppression}
    S(k) = 1- \left(\frac{2 k_c}{k} \right)^2 \sin^2 \left(\frac{k}{2 k_c} \right) ,
\end{equation}
where $k_c$ is the horizon scale at the Q-ball formation.
Thus, Eq.~\eqref{eq:Q-ball_power_spec} should be multiplied by $S(k)$ for the gravity-mediation type.
On the other hand, in the case of the gauge-mediation type Q-balls, charge conservation does not require energy conservation, and hence no suppression factor is needed.~\footnote{
The formed Q-balls go into the ground state by emitting extra (charge-neutral) radiation.
As a result, the total energy of Q-balls is not conserved.
In this paper, we assume that this process takes place rapidly.
}

\section{PBH formation by sub-horizon fluctuations}
\label{sec:PBH_formation}

Let us consider the formation of PBHs when density perturbations with sub-horizon scales are produced in a matter-dominated (MD) era.
The PBH formation in the early MD universe was studied in~\cite{Khlopov:1980mg,Polnarev:1985btg} and more recently in \cite{Harada:2016mhb,Georg:2019jld}. 
However, the previous work (implicitly) assumed the existence of super-horizon scale density perturbations, and it is not clear if their formula applies to density fluctuations generated inside the horizon.
Thus, we reconsider the condition of PBH formation in the case of the generation of sub-horizon density perturbations.
We follow the methodology in~\cite{Harada:2016mhb}.

First, we adopt the Zel'dovich approximation and write the coordinate as
\begin{equation}
    r_j = a(t) q_j + b(t) p_j(q_j)
    \qquad (j=1,2,3) ,
\end{equation}
where $a(t)$ is the scale factor, $q_j$ is the Lagrange coordinate, $p_j$ is the deviation vector, and $b(t)$ is the function denoting the linear growth of the perturbation.
The deformation tensor is defined as
\begin{equation}
    D_{jk} = \frac{\partial r_j}{\partial q_k}
    = a(t)\delta_{jk} +b(t)  \frac{\partial p_j}{\partial q_k}.
\end{equation}
We can take the axes for which $\partial p_i/\partial q_k$ is diagonal, i.e. $\partial p_i/\partial q_k = \mathrm{diag}(-\alpha, -\beta, -\gamma)$, which leads to
\begin{equation}
    \label{eq:deformation_tensor}
    D_{jk} = \mathrm{diag}\left(a(t)-\alpha b(t), 
    a(t)-\beta b(t), a(t)-\gamma b(t)\right).
\end{equation}
Using $\rho d^3 r = \bar{\rho}d^3q$ ($\bar{\rho}$ is the homogeneous density of the universe), the linear density perturbation $\delta_L$ is given by
\begin{equation}
    \delta_L = \frac{a^3}{(a-\alpha b)(a-\beta b)(a-\gamma b)}-1
    \simeq (\alpha+\beta+\gamma)\frac{b}{a}.
\end{equation}
Since $\delta_L$ grows as $\delta_L \propto a$ in the MD era, $b\propto a^2$.
From Eq.~\eqref{eq:deformation_tensor} we can take the following coordinate:
\begin{align}
    r_1 & = (a(t) -\alpha b(t))q_1 ,\\
    r_2 & = (a(t) -\beta b(t))q_2  ,\\
    r_3 & = (a(t) -\gamma b(t))q_2 ,
\end{align}
with $\alpha \ge \beta \ge \gamma $ and $\alpha >0$.

Let us consider the collapse of an overdensity region.
Suppose that the density perturbations are generated at $t=t_i$ inside the horizon.
The mass within unperturbed physical radius $a(t_i)q$ is given by
\begin{equation}
    M = \frac{4\pi}{3}\bar{\rho}(t_i) a^3(t_i) q^3.
\end{equation}
Introducing the parameter $\xi (< 1)$ that represents the ratio of the radius $a(t_i)q$ to the Hubble radius $H^{-1}(t_i)$, the mass $M$ is written as
\begin{align}
    M = \frac{H^2(t_i)}{2G} H^{-3}(t_i) \xi^3
    = \frac{\xi^3}{2G H(t_i)},
\end{align}
where we used the Friedmann equation $H^2 = (8\pi G/3)\bar{\rho}$.
The Schwarzschild radius $r_g$ is written as
\begin{equation}
    r_g = 2GM = H^{-1}(t_i) \xi^3 = a(t_i)q \xi^2 \equiv a_i q\xi^2.
\end{equation}

At the maximum expansion $t=t_m$, $\dot{r}_1 =0$ which gives $\dot{a}(t_m) = \alpha \dot{b}(t_m)$.
Using $b\propto a^2$, we obtain 
\begin{equation}
    \label{eq:bm/am}
    \frac{b_m}{a_m} = \frac{1}{2\alpha},
\end{equation}
where $a_m =a(t_m)$ and $b_m =b(t_m)$.
$r_m = r_1(t_m)$ is then written as
\begin{equation}
    r_m = (a_m -\alpha b_m)q  = \frac{1}{2}a_m q .
\end{equation}
On the other hand, using Eq.~\eqref{eq:bm/am}, $b_i/b_m\, (b_i \equiv b(t_i))$ is given by
\begin{equation}
    \frac{b_i}{b_m} = \frac{a_i^2}{a_m^2} =2\alpha \frac{b_i}{a_m},
\end{equation}
from which we obtain 
\begin{equation}
    \frac{a_i}{a_m} = 2\alpha \frac{b_i}{a_i} .
\end{equation}

When the region collapses into a pancake at $t=t_c$, $r_1(t_c) = 0$ from which we obtain
\begin{equation}
    \frac{b_c}{a_c}= \frac{1}{\alpha},
\end{equation}
where $a_c =a(t_c)$ and $b_c =b(t_c)$.
At $t=t_c$, $r_2$ and $r_3$ are given by
\begin{align}
    r_2(t_c) = 4r_m \left(1-\frac{\beta}{\alpha}\right) ,
    \label{eq:ellipse_2}\\
    r_3(t_c) = 4r_m \left(1-\frac{\gamma}{\alpha}\right) ,
     \label{eq:ellipse_3}
\end{align}
where we used $a_c = 2a_m$.

Now let us consider the condition for the pancake to form a black hole.
Using the hoop conjecture~\cite{Misner:1973prb}, a black hole is formed if the circumference of the ellipse given by Eqs.~\eqref{eq:ellipse_2} and \eqref{eq:ellipse_3} is less than $2\pi r_g$.
The circumference is given by
\begin{equation}
    \mathcal{C} = 16\, \frac{\alpha-\gamma}{\alpha}\,
    \mathrm{E}(e)r_m,
\end{equation}
where $\mathrm{E}(x)$ is the complete elliptic integral of the second kind, and $e$ is the eccentricity, $e^2 = 1-(\alpha-\beta)^2/(\alpha-\gamma)^2$.
Since $r_g$ is written as $r_g = a_iq\xi^2 r_m = 2\xi^2 (a_i/a_m)r_m
=4\alpha \xi^2 (b_i/a_i) r_m$, the condition for PBH formation is written as
\begin{align}
    h(\alpha,\beta,\gamma) \equiv 
    \frac{2}{\pi}\, \frac{\alpha-\gamma}{\alpha^2}\, \mathrm{E}(e)
    \lesssim \xi^2 .
\end{align}
Here we normalize $b(t)$ to $b(t_i)=a(t_i)$.

Assuming that the independent components of the symmetric deformation tensor have Gaussian distributions, the probability distribution of $\alpha$, $\beta$, and $\gamma$ is given by~\cite{1970Afz.....6..581D}
\begin{align}
    w(\alpha,\beta,\gamma;\sigma_M) 
    =&  -\frac{27\cdot5^{5/2}}{8\pi\sigma_M^6}
    \exp\left[ -\frac{3}{\sigma_M^2}\left( (\alpha^2+\beta^2+\gamma^2) 
    -\frac{1}{2}(\alpha\beta +\beta\gamma + \gamma\alpha)\right)\right] 
    \nonumber \\
    & \times (\alpha-\beta)(\beta-\gamma)(\gamma-\alpha).
\end{align}
Here $\sigma_M^2 = \langle\delta_L^2(t_i)\rangle$.
Using $w$, the probability of PBH formation is 
\begin{align}
    \beta_f = \int_0^\infty d\alpha \int_{-\infty}^\alpha d\beta
    \int_{-\infty}^\beta d\gamma \,\Theta(\xi^2-h(\alpha,\beta,\gamma))w(\alpha,\beta,\gamma;\sigma_M) ,
\end{align}
where $\Theta(x)$ is the Heaviside function.

Now we redefine $\alpha$, $\beta$, $\gamma$ and $\sigma$ as
\begin{align}
    \alpha  = \hat{\alpha}\xi^{-2} , \qquad
    \beta  = \hat{\beta}\xi^{-2} , \qquad
    \gamma  = \hat{\gamma}\xi^{-2} , \qquad
    \sigma_M = \hat{\sigma}_M \xi^{-2} .
\end{align}
We then obtain
\begin{align}
    h(\alpha,\beta,\gamma) 
    & = h(\hat{\alpha},\hat{\beta},\hat{\gamma}) ,\\
    w(\alpha,\beta,\gamma;\sigma)d\alpha d\beta d \gamma
    & = w(\hat{\alpha},\hat{\beta},\hat{\gamma};\hat{\sigma}_M)
    d\hat{\alpha} d\hat{\beta} d\hat{\gamma}  ,
\end{align}
from which the formation probability is rewritten as 
\begin{align}
    \beta_f = \int_0^\infty d\hat{\alpha} \int_{-\infty}^{\hat{\alpha}} d\hat{\beta}
    \int_{-\infty}^{\hat{\beta}} d\hat{\gamma} \,\Theta(1-h(\hat{\alpha},\hat{\beta},\hat{\gamma})
    w(\hat{\alpha},\hat{\beta},\hat{\gamma};\hat{\sigma}_M) .
\end{align}
This is the same as the probability that the region with Hubble radius collapses into a PBH, and we know the result~\cite{Harada:2016mhb},
\begin{equation}
    \beta_f = 0.056\hat{\sigma}_M^5 .
\end{equation}
Therefore, we finally obtain
\begin{equation}
    \label{eq:production_rate}
     \beta_f = 0.056\sigma_M^5 \xi^{10}
     = 0.056\sigma_M^5 \left(\frac{M}{M_H}\right)^{10/3},
\end{equation}
where $M_H$ is the horizon mass and $M = \xi^3 M_H$.
Notice that $\sigma_M$ and $M_H$ in Eq.~\eqref{eq:production_rate} are evaluated at $t=t_i$.
However, since $\sigma_M(t) = (a/a_i)\sigma_M(t_i)$ and $M_H(t) = (a/a_i)^{3/2}M_H(t_i)$ at later epochs, Eq.~\eqref{eq:production_rate} is satisfied for $\sigma_M$ and $M_H$ evaluated at $t > t_i$.
Notice that Eq.~\eqref{eq:production_rate} is the same as that derived assuming the existence of the superhorizon scale perturbations.
Therefore, we have shown that Eq.~\eqref{eq:production_rate} can be used for both super- and sub-horizon perturbations.

\section{PBH formation by Q-balls}
\label{sec:PBH_formation_Qball}

In this section, we estimate the abundance of PBHs.
For a concrete calculation, we consider the gauge-mediation type Q-balls, and use the result of the lattice simulation in~\cite{Kasuya:2025nix}.
The simulation adopts the following scalar potential:
\begin{equation}
    \label{eq:simulation_potential}
    V(\phi) = m^4 \left[\log\left(1+\frac{|\phi |}{m}\right)^2\right]^2
    \simeq m^4 \left[\log\left(\frac{|\phi |^2}{m^2}\right)\right]^2
    \qquad (|\phi| \gg m) ,
\end{equation}
which leads to the production of the gauge-mediated type Q-balls with $p=3/4$.
The simulation~\cite{Kasuya:2025nix}  shows that the average charge distribution is approximately given by the power law with $\alpha_Q \simeq -0.25$.\footnote{
In Ref.~\cite{Kasuya:2025nix}, the charge distribution is fitted to the function given by $\bar{N}(Q) \propto Q^{\alpha_Q-1} \exp(-\kappa_Q Q^2)$ ($\kappa_Q$: a constant).
In this paper, for simplicity, we adopt Eq.~\eqref{eq:Q-ball_charge_dist} as a fitting function with $Q_*$ being the peak charge that dominates the energy density of Q-balls.
}

Furthermore, from the simulation, we can estimate the average number of the formed Q-balls with $Q> Q_m$ in the horizon volume $V_H= (4\pi/3)H(t_f)^{-3}$ at the formation time $t_f$, which we denote $\bar{N}(Q>Q_m)$.
The simulation~\cite{Kasuya:2025nix} gives $\bar{N}(Q> 0.01Q_*) \simeq 1.5\times 10^5$.
Using Eq.~\eqref{eq:Q-ball_charge_dist} we obtain 
\begin{equation}
    \int_{0.01Q_*}^{Q_*} \frac{dQ}{Q}\, \mathcal{N} V_H 
    \left(\frac{Q}{Q_*}\right)^{\alpha_Q} 
    \simeq 1.5\times 10^5,
\end{equation}
from which we determine $\mathcal{N}_Q V_H\simeq  1.7\times 10^4$.
Thus, the Q-ball density perturbation at the horizon scale is 
\begin{equation}
     \sigma_{Q,H} = \sigma_Q(V_H) \simeq 3.4\times 10^{-3} .
\end{equation}
Since the variance $\sigma_Q^2$ of the Q-ball density perturbation in a volume $V$ is proportional to $V^{-1}$ (see Eq.~\eqref{eq:Q-ball_sigma})), $\sigma_Q$ at smaller scales is written as
\begin{equation}
    \label{eq:Q-ball_perturbation}
    \sigma_Q (V) = \sigma_{Q,H}\left(\frac{V}{V_H}\right)^{-1/2}.
\end{equation}
Since we assume the linear growth of the density perturbations, the initial Q-ball perturbations must be smaller than $1$, which requires that the Q-ball number in the volume $V$ ($\sim \mathcal{N}_Q V$) should be larger than 1.
Moreover, we later approximate the Poisson distribution by the Gaussian one (see discussion below Eq.~\eqref{eq:pbh_formation_rate_from_Qball}), which is valid for  $\mathcal{N}_Q V \gg 1$.
Thus, taking into account $\mathcal{N}_Q V_H \sim 10^4$, we use Eq.~\eqref{eq:Q-ball_perturbation} for $V\gtrsim 10^{-2} V_H$.

When the Q-balls dominate the universe, the Q-ball density perturbations are equal to the total density perturbations.
From Eq.~\eqref{eq:Q-ball_perturbation} the density perturbation $\sigma(M)$ smoothed over the volume $V=M/\rho_Q$ is given by
\begin{equation}
    \label{eq:density_pert}
    \sigma(M) = \sigma_{Q,H}\left(\frac{M}{M_H}\right)^{-1/2}.
\end{equation}
The horizon mass at the Q-ball formation time $t_f$ is given by
\begin{equation}
    M_H(t_f) \equiv M_{f,H}= \frac{4\pi}{3}\rho_\mathrm{tot} H^{-3}(t_f)
    = \frac{4\pi M_\mathrm{pl}^2}{H(t_f)},
\end{equation}
where $\rho_\mathrm{tot}$ is the total density of the universe, and $M_\mathrm{pl}\,(=2.4\times 10^{18}\,\mathrm{GeV})$ is the Planck mass.
The scalar field $\phi$ that forms the Q-balls starts to oscillate when the Hubble parameter  $H$ becomes equal to the effective mass of the scalar field $m_{\phi, \mathrm{eff}}$.
The simulation~\cite{Kasuya:2025nix} shows that the Q-ball formation takes place when the scale factor increases by about a factor of $6$ since the oscillation starts at $t=t_\mathrm{osc}$.
Thus, assuming a matter-dominated era, we obtain
\begin{equation}
    H(t_f) = \left(\frac{a(t_\mathrm{osc})}{a(t_f)}\right)^{3/2} H(t_\mathrm{osc})
    \simeq  6.8\times10^{-2} m_{\mathrm{eff}}
    ,
\end{equation}
from which the Horizon mass is written as
\begin{align}
     M_{f,H} & = 
    0.98 \times 10^{-15}\,M_\odot
    \left(\frac{ m_{\mathrm{eff}}}{\mathrm{MeV}}\right)^{-1}
    \nonumber \\[0.4em]
    & = 
    2.0 \times 10^{18}\,\mathrm{g}
    \left(\frac{ m_{\mathrm{eff}}}{\mathrm{MeV}}\right)^{-1}
    \label{eq:horizon_mass}
\end{align}
The density perturbations induced by the Q-balls lead to the primordial black hole formation.
Using Eqs.~\eqref{eq:production_rate} and \eqref{eq:density_pert} we obtain the PBH production rate as
\begin{equation}
    \label{eq:pbh_formation_rate_from_Qball}
    \beta_f(M)  
     = 0.056\,\sigma_{Q,H}^5 \left(\frac{M}{M_{f,H}}\right)^{5/6}
     \qquad (M_{f,H} \gtrsim M \gtrsim 10^{-2}M_{f,H} )
\end{equation}
Here we assume that the PBH mass is given by $M$.
Notice that strictly speaking, Eq.~\eqref{eq:production_rate} can be used when the density perturbations are Gaussian.
Since the Poisson distribution is considered as Gaussian if $\mathcal{N}_Q V$ is large and $\mathcal{N}_Q V \gtrsim 10^2 \gg 1$ in the present case, we can apply Eq.~\eqref{eq:production_rate} to the Q-ball density perturbations.

So far we have considered the sub-horizon density perturbations of the Q-balls.
However, as mentioned in Sec.~\ref{sec:Qball_perturbation}, the gauge-mediation type Q-balls can also produce super-horizon scale perturbations.
In this case, the variance is given by Eq.~\eqref{eq:density_pert} with  $M > M_H$.
Since those perturbations grow after the horizon entry, the PBH formation rate is given by 
\begin{equation}
    \beta_f(M) = 0.056\,\sigma_{Q,H}^5 \left(\frac{M}{M_{f,H}}\right)^{-5/2}
     \qquad (M_{R,H} \gtrsim M \gtrsim M_{f,H} ),
\end{equation}
where $M_{R,H}$ is the horizon mass when the universe becomes radiation-dominated.

We assume that the Q-balls decay into radiation with temperature $T_\mathrm{dec}$ and the radiation-dominated universe starts.
During the radiation-dominated era, the density of the produced PBHs $\rho_\mathrm{PBH}$ increases proportionally to the scale factor $a$ until the matter-radiation equality.
Noticing that $\rho_\mathrm{PBH}/\rho_R \propto a^{-3}/g_*T^4 = (g_* T^3 a^3)^{-1}T^{-1} \propto 1/T$ ($\rho_R$: radiation density, $g_*$: number of relativistic degrees), the present PBH fraction of the dark matter $f(M)$ is given by
\begin{align}
    f(M) & = \beta_f(M) \frac{T_\mathrm{dec}}{T_\mathrm{eq}}
    \frac{\Omega_m}{\Omega_\mathrm{DM}}
    \nonumber \\
    &= 6.9\times 10^{-2} \,
    \left(\frac{\sigma_{Q,H}}{10^{-3}}\right)^5 
    \left(\frac{M}{M_{f,H}}\right)^{5/6} 
    \left(\frac{T_\mathrm{dec}}{10^6\,\mathrm{GeV}}\right)
    \frac{\Omega_m}{\Omega_\mathrm{DM}}
    \nonumber \\[0.4em]
    & \qquad (M_{f,H} \gtrsim M \gtrsim 10^{-2}M_{f,H} )
    ,
    \label{eq:PBH_abundance}
    \\
    f(M) &= 6.9\times 10^{-2} \,
    \left(\frac{\sigma_{Q,H}}{10^{-3}}\right)^5 
    \left(\frac{M}{M_{f,H}}\right)^{-5/2} 
    \left(\frac{T_\mathrm{dec}}{10^6\,\mathrm{GeV}}\right)
    \frac{\Omega_m}{\Omega_\mathrm{DM}}
    \nonumber \\[0.4em]
    & \qquad (M_{R,H} \gtrsim M \gtrsim M_{f,H} )
    ,
    \label{eq:PBH_abundance_super_horizon}
\end{align}
where $T_\mathrm{eq} \,(= 0.81\,\mathrm{eV})$ is the temperature at the matter-radiation equality, $\Omega_m \,(\simeq 0.31)$ and $\Omega_\mathrm{DM}\, (\simeq 0.26)$ are the density parameters of matter and dark matter, respectively.
At the Q-ball decay, the Hubble parameter $H_\mathrm{dec}$ is written as
\begin{equation}
    H_\mathrm{dec} = \sqrt{\frac{\pi^2 g_* }{90}}\,\frac{T_\mathrm{dec}^2}{M_\mathrm{pl}},
\end{equation}
from which the horizon mass at the Q-ball decay is estimated as
\begin{equation}
    M_{R,H} = \frac{4\pi M_\mathrm{pl}^2}{H_\mathrm{dec}}
    = 4.9\times 10^{-14} \, M_\odot 
    \left(\frac{T_\mathrm{dec}}{10^6\,\mathrm{GeV}}\right)^{-2}
    \left(\frac{g_*}{100}\right)^{-1/2}.
\end{equation}

\begin{figure}[t]
    \centering
    \includegraphics[width=.75\textwidth ]{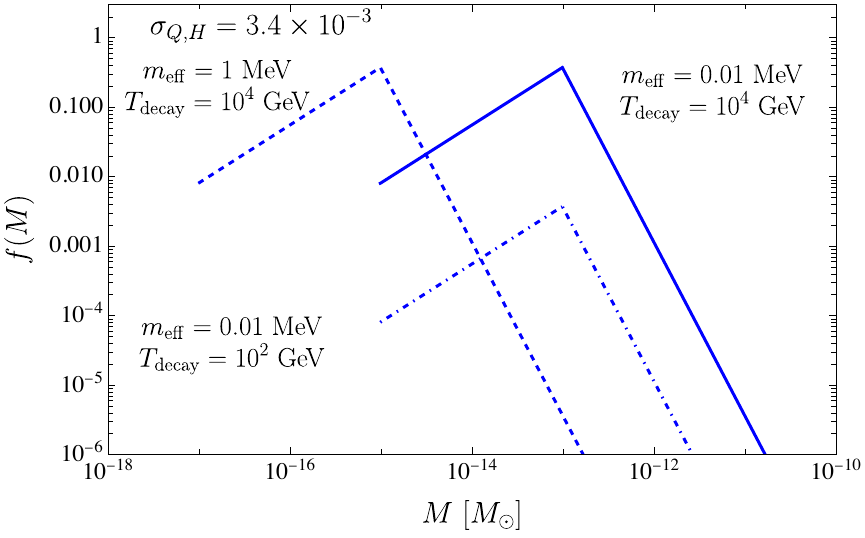}
    \caption{%
        The PBH fraction for $\sigma_{Q,H}=3.4\times 10^{-3}$.
        The solid, dashed, and dot-dashed lines are for ($m_\mathrm{eff}/\mathrm{MeV}$, $T_\mathrm{dec}/\mathrm{GeV}$) = ($0.01, 10^4)$, $(1, 10^4)$, and $(0.01, 10^2)$, respectively.
    }
    \label{fig:PBH_frac}
\end{figure}

\begin{figure}[t]
    \centering
    \includegraphics[width=.75\textwidth ]{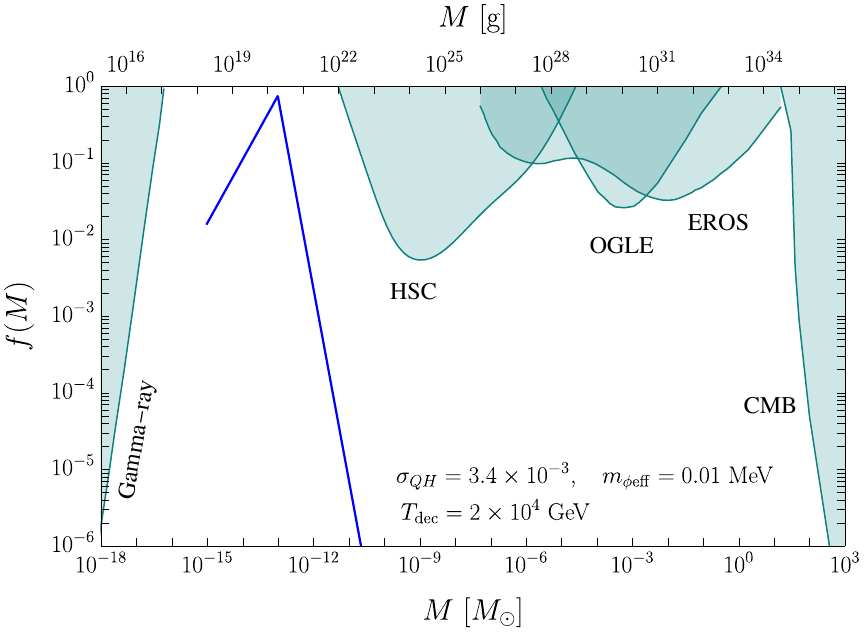}
    \caption{%
        The PBH fraction for $\sigma_{Q,H}=3.4\times 10^{-3},~T_\mathrm{dec}=2 \times 10^4\mathrm{GeV}$, $m_{\mathrm{eff}}=0.01\mathrm{MeV}$.
        The shaded regions show the observational constrains from $\gamma$-rays~\cite{Carr:2009jm}, microlensing(HSC~\cite{Niikura:2017zjd}, OGLE~\cite{Niikura:2019kqi}, EROS~\cite{EROS-2:2006ryy}), and CMB~\cite{Serpico:2020ehh}.  
    }
    \label{fig:PBH_frac_obs}
\end{figure}

Figure~\ref{fig:PBH_frac} shows the PBH fraction of the dark matter for three sets of parameters ($\sigma_{Q,H}$, $m_{\mathrm{eff}}/\mathrm{MeV}$, $T_\mathrm{dec}/\mathrm{GeV}$) $= (3.4\times 10^{-3}, 0.01, 10^4)$, $(3.4\times 10^{-3}, 1, 10^4)$, and $(3.4\times 10^{-3}, 0.01, 10^2)$.
It is seen that the peak of the PBH fraction is given by the horizon mass at the Q-ball formation (Eq.~\eqref{eq:horizon_mass}), which is determined by the $m_{\mathrm{eff}}$.
The PBH abundance depends on the variance of the Q-ball density perturbation at the horizon scale $\sigma_{Q,H}$ and the decay temperature $T_\mathrm{dec}$ as $\sigma_{Q,H}^5 T_\mathrm{dec}$.
We also show the PBH fraction for $\sigma_{Q,H}=3.4\times 10^{-3},~T_\mathrm{dec}=2 \times 10^4\mathrm{GeV},~m_{\mathrm{eff}}=0.01\mathrm{MeV}$ together with the observational constraints~\cite{Carr:2009jm,Niikura:2017zjd,Niikura:2019kqi,EROS-2:2006ryy,Serpico:2020ehh} in Fig.~\ref{fig:PBH_frac_obs}.

\section{Constraints on PBH formation}
\label{sec:constraint}

For this PBH formation scenario to work, the Q-ball should decay after the PBH formation, i.e. $H_\mathrm{dec}$ should be smaller than $H(t_f)\simeq m_{\mathrm{eff}}(a(t_\mathrm{osc})/a(t_f))^{3/2}$.
Thus, we obtain a constraint on $T_\mathrm{dec}$ as
\begin{equation}
    T_\mathrm{dec} \lesssim 
    0.71\times 10^7\, \mathrm{GeV}\, 
    \left(\frac{m_{\mathrm{eff}}}{\mathrm{MeV}}\right)^{1/2}
    \left(\frac{g_*}{100}\right)^{-1/4}.
\end{equation}
This leads to the upper bound on the PBH fraction as
\begin{equation}
    \label{eq:const1}
    f(M) \lesssim 
    0.49\,
    \left(\frac{\sigma_{Q,H}}{10^{-3}}\right)^5 
    \left(\frac{M}{M_H}\right)^{5/6} 
    \left(\frac{m_{\mathrm{eff}}}{\mathrm{MeV}}\right)^{1/2}
    \left(\frac{g_*}{100}\right)^{-1/4}
    \frac{\Omega_m}{\Omega_\mathrm{DM}}.
\end{equation}

\begin{figure}[t]
    \centering
    \includegraphics[width=.75\textwidth ]{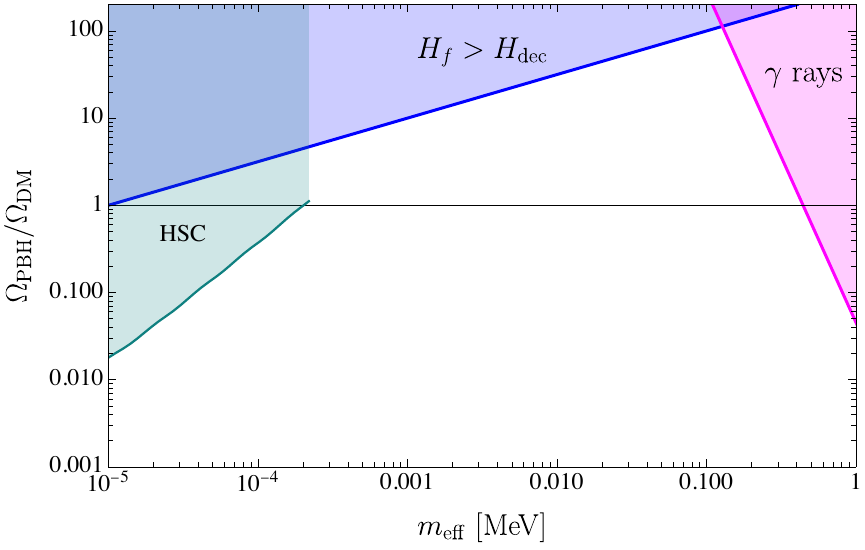}
    \caption{%
        Constraints on $\Omega_\mathrm{PBH}/\Omega_\mathrm{DM}$.
        The blue region is excluded from the constraint~\eqref{eq:const1}
        while the red region is excluded from~\eqref{eq:const2}.
        The green region is excluded from the miclolensing observation by HSC~\cite{Niikura:2017zjd}.
    }
    \label{fig:const_PBH_abundance}
\end{figure}

Eqs.~\eqref{eq:horizon_mass} and \eqref{eq:PBH_abundance} imply that the produced PBHs have masses between about $10^{16}$~g and $10^{20}$~g for $m_{\mathrm{eff}}\sim (1-0.01)$~MeV.
Small PBHs evaporate emitting Hawking radiation, and PBHs with mass smaller $5\times 10^{14}$~g completely evaporate until the present.
On the other hand, PBHs with larger mass survive and emit photons.
The abundance of those small PBHs is stringently constrained by observations of $\gamma$-rays as~\cite{Carr:2009jm}
\begin{equation}
    \label{eq:gamma_const}
    f(M) \lesssim 0.4\,\left(\frac{M}{10^{17}\,\mathrm{g}}\right)^{3.8}.
\end{equation}
Since the constraint is the most stringent for the smallest PBHs, we use  Eqs.~\eqref{eq:PBH_abundance} and \eqref{eq:gamma_const} with $M=M^{-2}M_{f,H}$ and obtain the following constraint:
\begin{equation}
    T_\mathrm{dec} \lesssim 
    5.4\times 10^5\, \mathrm{GeV}
    \left(\frac{\sigma_{Q,H}}{10^{-3}}\right)^{-5} 
    \left(\frac{m_{\mathrm{eff}}}{\mathrm{MeV}}\right)^{-3.8}
    \frac{\Omega_\mathrm{DM}}{\Omega_m}.
\end{equation}
This then gives the upper bound on the PBH abundance as
\begin{equation}
    \label{eq:const2}
    f(M) \lesssim 
    0.038\,
    \left(\frac{M}{M_{f,H}}\right)^{5/6} 
    \left(\frac{m_{\mathrm{eff}}}{\mathrm{MeV}}\right)^{-3.8}.
\end{equation}
In Fig.~\ref{fig:const_PBH_abundance} we show the constraints on $\Omega_\mathrm{PBH}/\Omega_\mathrm{DM}$ coming from Eqs.~\eqref{eq:const1} and \eqref{eq:const2}.
In the figure, we integrate $f(M)$ over $M$.

When the PBH mass is larger than $5\times 10^{-12}M_\odot$ (which corresponds to $m_{\mathrm{eff}} \lesssim 2\times 10^{-4}$), the microlensing observations give constraints on the PBH abundance as shown in Fig.~\ref{fig:PBH_frac_obs}.
Since the PBH mass function has a peak at $M_{f,H}$ and drops sharply at larger masses, we can obtain the constraint on $m_{\mathrm{eff}}$ and $\Omega_\mathrm{PBH}$ by comparing $f(M_{f,H})$ to the constraint from the HSC microlensing, which is also shown in~\ref{fig:const_PBH_abundance}.
From Fig.~\ref{fig:const_PBH_abundance} it is seen that the PBHs can account for all dark matter of the universe for $m_\mathrm{eff} \simeq 2\times 10^{-4} - 0.4$~MeV.

\section{Q-balls in supersymmetric models}
\label{sec:SUSY_model}

Here we show that the present scenario is realized in the supersymmetric (SUSY) framework.
Q-balls are produced through the Affleck-Dine mechanism~\cite{Kusenko:1997si,Dvali:1997qv,Enqvist:1997si,Kasuya:1999wu} that uses one of the flat directions called the Affleck-Dine (AD) field in the minimal supersymmetric standard model and generates a baryon or lepton asymmetry in the universe~\cite{Affleck:1984fy,Dine:1995kz}.  
During inflation, the AD field $\Phi$ has a large field value and starts to oscillate when the effective mass becomes equal to the Hubble parameter after inflation.
During oscillation, the AD field develops spatial instabilities and forms Q-balls.
The Q-ball properties depend on the SUSY breaking scheme.
We here consider the gauge-mediated SUSY breaking, which gives the SUSY breaking potential~\cite{deGouvea:1997afu},
\begin{equation}
    \label{eq:potential}
    V(\Phi ) = M_F^4 \left[ \log \left(\frac{|\Phi|^2}{M_\mathrm{mess}^2}\right)\right]^2
    \qquad (|\Phi| \gtrsim M_\mathrm{mess}),
\end{equation}
where $M_F$ is the SUSY breaking scale and $M_\mathrm{mess}$ is the messenger scale.
Notice that this potential is essentially the same as the potential~\eqref{eq:simulation_potential} used in the simulation if we identify $m$ with $M_F$.
The mass and radius of the Q-ball are given by~\cite{Hisano:2001dr}
\begin{align}
    M_Q &\simeq \frac{4\sqrt{2}\pi}{3}\zeta M_F Q^{3/4}, \\
    R_Q & \simeq \frac{1}{\sqrt{2}\zeta}M_F^{-1} Q^{1/4},
\end{align}
where $\zeta$ ($\sim \mathcal{O}(1)$) is a numerical factor.
$\mu$ in Eq.~\eqref{eq:Q-ball_mass} is then written as $\mu = \frac{4\sqrt{2}\pi}{3}\zeta M_F$ and $p=3/4$.

The equation of motion for the AD field is written as
\begin{equation}
    \ddot{\Phi} +3H\dot{\Phi} + m_\mathrm{eff}^2\Phi =0.
\end{equation}
From Eq.~\eqref{eq:potential} the effective mass $m_\mathrm{eff}$ is given by
\begin{equation}
    \label{eq:effective_mass_AD}
    m_\mathrm{eff}^2 = \frac{2M_F^4}{|\Phi|^2}
    \log \left(\frac{|\Phi|^2}{M_\mathrm{mess}^2}\right).
\end{equation}
When the AD field starts to oscillate ($m_\mathrm{eff} \simeq H$), the ratio of the density of the AD field $\rho_\mathrm{AD}$ to the total density of the universe $\rho_\mathrm{tot}$ is 
\begin{equation}
    \label{eq:AD_density_frac}
    \frac{\rho_\mathrm{AD}}{\rho_\mathrm{tot}} 
    = \frac{M_F^4 \left[\log (|\Phi_\mathrm{osc}|^2/M_\mathrm{mess}^2)\right]^2}{3H_\mathrm{osc}^2M_\mathrm{pl}^2}
    \simeq 
    \frac{|\Phi_\mathrm{osc}|^2\log (|\Phi_\mathrm{osc}|^2/M_\mathrm{mess}^2)}{6M_\mathrm{pl}^2}.
\end{equation}
For the PBH formation scenario to work, the AD field should dominate the universe when or immediately after the oscillation starts; otherwise, the density perturbations of the Q-balls are reduced, leading to a significant decrease in the PBH abundance. 
Thus, we require $\rho_\mathrm{AD}/\rho_\mathrm{tot}\sim 1$, from which we obtain
\begin{equation}
    \label{eq:AD_field_value}
    |\Phi_\mathrm{osc}| \sim 0.35 M_\mathrm{pl},
\end{equation}
for $M_\mathrm{mess} = 10^{6}$\,--\,$10^8$~GeV.
With use of Eqs~\eqref{eq:effective_mass_AD}, \eqref{eq:AD_density_frac} and \eqref{eq:AD_field_value}, the SUSY breaking scale $M_F$ is then related to $m_\mathrm{eff}$ as
\begin{equation}
    M_F \sim \frac{1}{(12)^{1/4}} (m_\mathrm{eff} M_\mathrm{pl})^{1/2}
    \left(\frac{|\Phi_\mathrm{osc}|}{M_\mathrm{pl}}\right)
    \sim 3\times 10^6\,\mathrm{GeV} \left(\frac{m_\mathrm{eff}}{0.1\,\mathrm{MeV}}\right)^{1/2}.
\end{equation}
Since we need $m_\mathrm{eff} \simeq 2\times 10^{-4} -0.4$~MeV to explain all dark matter of the universe, the SUSY breaking scale should be $M_F\sim (0.1\,\textrm{--}\, 6)\times 10^6$~GeV.
Furthermore, from Eq.~\eqref{eq:horizon_mass} the peak PBH mass is written as
\begin{equation}
    M_\mathrm{PBH} \sim 0.6\times 10^{-13}\,M_\odot\left(\frac{M_F}{10^6\,\mathrm{GeV}}\right)^{-2}.
\end{equation}

In the minimal gauge-mediated SUSY breaking model, SUSY is broken by the SUSY breaking field $S$ whose $F$-term is given by $F_S$.
The SUSY breaking sector is connected to the observable sector by a pair of the messenger fields $\Psi$ and $\bar{\Psi}$.
The masses of the gauginos and sferminos are then written as
\begin{align}
    M_\mathrm{gaugino} &\simeq \frac{g^2}{16\pi^2}
    \frac{kF_S}{M_\mathrm{mess}}
    =\frac{g^2}{16\pi^2}\Lambda_\mathrm{mess},\\
    M_\mathrm{sfermion} &\simeq \frac{g^2}{16\pi^2}
    \Lambda_\mathrm{mess} ,
\end{align}
where $\Lambda_\mathrm{mess}= k F_S/M_\mathrm{mess}$ and $k$ is the coupling constant.
$kF_S$ and $M_\mathrm{mess}$ satisfy $\sqrt{kF_S}< M_\mathrm{mess}$ ($\Lambda_\mathrm{mess} < M_\mathrm{mess}$) to avoid tachyons.
The SUSY breaking scale $M_F$ is related to $\Lambda_\mathrm{mess}$ and $M_\mathrm{mess}$ as
\begin{equation}
    M_F = \frac{g^{1/2}}{4\pi}\sqrt{\Lambda_\mathrm{mess} M_\mathrm{mess}} 
    \simeq 10^6\,\mathrm{GeV} 
    \left(\frac{\Lambda_\mathrm{mess}}{2\times 10^6\mathrm{GeV}}\right)^{1/2}
    \left(\frac{M_\mathrm{mess}}{8\times 10^7\mathrm{GeV}}\right)^{1/2}.
\end{equation}
Therefore, $M_F$ required for the formation of dark matter PBH is realized when $\Lambda_\mathrm{mess} \simeq  0.025 M_\mathrm{mess} \simeq 2\times 10^6$~GeV.
From the  scale of the messenger sector, we obtain the masses of the SUSY particles as
\begin{align}
    M_\mathrm{gaugino} \sim M_\mathrm{sfermion}\sim  10\,\mathrm{TeV}
\end{align}

Here, we remark on the mass of the gravitino.
When the SUSY breaking is caused only by the field $S$, the gravitino mass is written as
\begin{equation}
    m_{3/2} \simeq \frac{1}{\sqrt{3}k}
    \frac{\Lambda_\mathrm{mess} M_\mathrm{mess}}{M_\mathrm{pl}}
    \simeq 0.04\,k^{-1}\mathrm{MeV},
\end{equation}
for $\Lambda_\mathrm{mess}\simeq 0.025 M_\mathrm{mess} \simeq 2\times 10^6$~GeV.
It is known that such light gravitinos cause the gravitino problem since they are produced through scattering in the thermal radiation and their density exceeds the dark matter density of the universe~\cite{Moroi:1993mb}.
In the present case, thermal radiation is produced at the Q-ball decay, and its temperature is $T_\mathrm{dec}\simeq 2\times 10^{4}$~GeV for the PBHs to explain the dark matter abundance.
The present abundance of the thermally produced gravitinos is given by~\cite{Kawasaki:2017bqm,Kawasaki:2022hvx}
\begin{equation}
    \Omega_{3/2}h^2 \simeq 0.035\, 
    \left(\frac{m_{3/2}}{\mathrm{GeV}}\right)^{-1}
    \left(\frac{m_{\tilde{g}}}{10\mathrm{TeV}}\right)^{2}
    \left(\frac{T_{R}}{10^4\mathrm{GeV}}\right),
\end{equation}
where $m_{\tilde{g}}$ is the gluino mass.
Thus, we obtain the constraint $m_{3/2}\gtrsim 1$~GeV to avoid the gravitino overproduction.
This requirement is satisfied without affecting the potential of the AD field if there exists a sequestered supersymmetry-breaking sector giving a large mass to the gravitino.

\section{Summary and Discussion}
\label{sec: summary and discussion}

In this paper, we have studied the density perturbations induced by Q-balls and derived the formula to calculate the density perturbations from the Q-ball charge distribution.
We have also re-examined the condition for the PBH formation in the matter-dominated era and showed that the previously derived formula for super-horizon density fluctuations can be used for the sub-horizon density perturbations.
As a concrete example, we considered the gauge-mediation type Q-balls whose charge distribution was obtained by the lattice simulation, and estimated the density perturbations assuming that the Q-ball dominates the universe.
It was found that the density perturbations are large enough to produce a significant number of PBHs with mass $10^{-15}\,M_\odot -5\times 10^{-12}\, M_\odot$ and can explain all the dark matter of the universe.

The most plausible scenario for the Q-ball formation is the Affleck-Dine mechanism within the framework of supersymmetry theory.
In the gauge-mediated SUSY breaking models, we found that, to produce the PBHs that account for the dark matter, the SUSY breaking scale is given by $M_F \sim 10^6$~GeV. 
This is consistent with the SUSY particles (squarks, sleptons and gauginos) with masses $\sim 10$~TeV.

In the present scenario, to
produce large density perturbations, the Q-balls need to dominate the universe at their formation.
This requires that the scalar field forming the Q-ball should have a large field value $\gtrsim 0.1 M_\mathrm{pl}$.
One interesting way to realize this is that the scalar field is an inflaton~\cite{Enqvist:2002si}.
Another requirement in this scenario is that the Q-balls should decay into radiation with a reheating temperature of $T_R \gtrsim 10^4$~GeV.
This is because a sufficient duration of the radiation-dominated period is necessary to increase the density fraction of the PBHs.
In general, Q-balls decay slowly since the decay only takes place near the surface of the Q-balls due to the Pauli blocking.
However, the Q-ball decay is enhanced if the $U(1)$ breaking interactions exist~\cite{Kawasaki:2025nsi,Cotner:2016dhw,Kawasaki:2019ywz,Kawasaki:2005xc}.

Finally, we make some comments on the application to the oscillons.
The oscillons are non-topological solitons that can be produced in the early universe from real scalar fields.
The formula derived in this paper can be applied to the oscillons by replacing the charge $Q$ with the mass $M$ of an oscillon and adding the factor~\eqref{eq:suppression} due to energy conservation. 
It is known that the oscillons are produced from axion-like fields with a non-cosine type potential~\cite{Amin:2011hj,Hong:2017ooe,Kawasaki:2019czd}.
In this case, the axion-like field can also play the role of the inflaton~\cite{Nomura:2017ehb} and realize the oscillon-dominated universe after inflation.

\begin{acknowledgments}
We are grateful to Satoshi Shirai for the valuable discussions.
This work was supported by JSPS KAKENHI Grant Numbers 25K07297 (M.K.), JP25KJ1030 (S.N.), and JST SPRING, Grant Number JPMJSP2108 (S.N.).
A.K. was supported by the U.S. Department of Energy (DOE) Grant No. DE-SC0009937 and by World Premier International Research Center Initiative (WPI), MEXT, Japan.
\end{acknowledgments}

\small
\bibliographystyle{JHEP}
\bibliography{Ref}

\end{document}